\documentclass[aps,pre,twocolumn,superscriptaddress,10pt,amsmath,amssymb,footinbib]{revtex4-1}

\usepackage{graphicx}

\begin{document}


\title{Wrinkling of a spherical lipid interface induced by actomyosin cortex}


\author{Hiroaki Ito}
\thanks{These authors contributed equally to this work.}
\affiliation{Department of Physics, Graduate School of Science, Kyoto University, Kyoto 606-8502, Japan}

\author{Yukinori Nishigami}
\thanks{These authors contributed equally to this work.}
\affiliation{Department of Physics, Graduate School of Science, Kyoto University, Kyoto 606-8502, Japan}

\author{Seiji Sonobe}
\affiliation{Department of Life Science, Graduate School of Life Science, University of Hyogo, Harima Science Park City, Hyogo 678-1297, Japan}

\author{Masatoshi Ichikawa}
\affiliation{Department of Physics, Graduate School of Science, Kyoto University, Kyoto 606-8502, Japan}




\begin{abstract}
Actomyosin actively generates contractile forces that provide the plasma membrane with the deformation stresses essential to carry out biological processes. Although the contractile property of purified actomyosin has been extensively studied, to understand the physical contribution of the actiomyosin contractile force on a deformable membrane is still a challenging problem and of great interest in the field of biophysics. Here, we reconstituted a model system with a cell-sized deformable interface that exhibits anomalous curvature dependent wrinkling caused by actomyosin cortex underneath the spherical closed interface. Through the shape analysis of the wrinkling deformation, we found that the dominant contributor on the wrinkled shape changes from bending elasticity to stretching elasticity of the reconstituted cortex by increasing the droplet curvature radius of the order of the cell-size, i.e., tens of micrometer. The observed curvature dependence was explained by the theoretical description of the cortex elasticity and contractility. Our present results provide a fundamental insight on the deformation of a curved membrane induced by the actomyosin cortex.
\end{abstract}



\maketitle



%

\section{INTRODUCTION}
Many types of eukaryotic cells have a thin shell-like structure, called the cell cortex, that underlies the cell membrane and consists of actin filaments (F-actin) and type II myosin. Whereas the static stiffness of the actomyosin cortex maintains the cellular shape against external stresses, the active contractile forces of the cortex play crucial roles in a variety of biological processes, such as cell motility \cite{Charras2008}, cell division \cite{Sedzinski2011}, embryonic development \cite{Blaser2006}, wound healing \cite{Zhao2009} and cancer metastasis \cite{Olson2009}. The actomyosin cortex is interconnected with the plasma membrane by actin-related proteins, such as ezrin-radixin-moesin (ERM) proteins \cite{Bretscher2002}. However, the mechanism of the consequent membrane deformation caused by force transduction between the cell cortex and cell membrane remains unclear because the membrane-cortex complex contains highly intricate components and connections \cite{Biro2013}. To understand the intricate roles of the actomyosin cortex in cellular processes, bottom-up or reconstitution approaches with a small number of essential components have been adopted. For example, actin polymerization and depolymerization with actin-related proteins in purified systems \cite{Blanchoin2000}, the rheological measurements of purified F-actin solution \cite{Kunita2012}, the motor activity of myosin from the perspective of its interaction with actin \cite{Heissler2011}, and single molecule analysis \cite{Nagy2013} have been examined thus far. As reconstitutions of biomimetic artificial systems, the surfaces of giant unilamellar vesicles \cite{Haeckl1998,Miyata1999,Liu2008,Koehler2011,Vogel2012} or beads \cite{Gucht2005,Paluch2006} have been utilized frequently to investigate the physicochemical properties of interfacial deformation or symmetry breaking of the surrounding actin cortices. On the other hand, actomyosin encapsulated within a cell-sized confinement exhibits distinct structures, such as network, cortex, and aster formations \cite{Koehler2011,Tsai2011,Takiguchi2011,Carvalho2013,Shah2014}. However, the fundamental issues regarding the actomyosin cortex, i.e., how the mechanical contractile forces exerted by the actomyosin affect the deformable plasma membrane and its shape, remain poorly understood because of the difficulties of in vitro reconstruction, such as those posed by the establishment of effective coupling between actomyosin and membranes \cite{Takiguchi2011,Carvalho2013,Shah2014}. Therefore, we sought to develop a model system that can exhibit the primitive aspects of cell membrane deformation due to the contractile force generated by membrane-coupled actomyosin.

\section{MATERIALS AND METHODS}
\subsection{Cell culture} Mass culture of {\it Amoeba proteus} was performed as described previously \cite{Nishihara2008}. In brief, {\it Amoeba proteus} was cultured in KCM medium (7 mg/l KCl, 8 mg/l CaCl$_2$, and 8 mg/l MgSO$_4 \cdot$7H$_2$O) at 25$^\circ$C and fed with {\it Tetrahymena pyriformis}. To avoid contamination with {\it T. pyriformis} cells were starved for at least 3 days before use.

\subsection{Preparation of the actomyosin fraction} The actomyosin fraction was prepared by modifying a previously described method \cite{Nishigami2013}. All procedures were performed at 2$^\circ$C. {\it A. proteus} (10 g) was suspended in a cell washing solution (2 mM O,O$^\prime$-Bis(2-aminoethyl)ethyleneglycol-N,N,N$^\prime$,N$^\prime$-tetraacetic acid (EGTA), 2 mM MgCl$_2$, and 20 mM Piperazine-1,4-bis(2-ethanesulphonic acid) (PIPES)-KOH, pH 7.0) and centrifuged at 6,000$\times g$ for 2 min, and then the supernatant was removed. The resultant precipitate was centrifuged at 600,000$\times g$ for 20 min to obtain an actin rich solution. The precipitate was suspended in a 3 M KCl solution (3 M KCl, 2 mM MgCl$_2$, 1 mM threo-1,4,-dimercapto-2,3-butanediol (DTT), 20 ${\rm \mu}$g/ml leupeptin, 20 ${\rm \mu}$g/ml pepstatin A, 0.2 mM phenylmethylsulphonyl fluoride (PMSF), 20 mM imidazole-HCl, pH 7.0) and centrifuged at 400,000$\times g$ for 10 min. The resultant supernatant was dialyzed against 50 mM KCl, 2 mM EGTA, 2 mM MgCl$_2$, 1 mM DTT, 0.2 mM PMSF, 20 mM imidazole-HCl, pH 7.0 for 5 h. The dialyzed solution was centrifuged at 20,000$\times g$ for 5 min, and the supernatant was discarded. The precipitate was suspended in a 300 ${\rm \mu}$l EMPA solution (5 mM EGTA, 6 mM MgCl$_2$, 1 mM DTT, 2 mM adenosine-5$^\prime$-triphosphate (ATP), 30 mM PIPES-KOH, pH 7.0), and 30 ${\rm \mu}$l actin rich solution was added. The obtained fraction dissolving the actomyosin contains 4.5 mM EGTA, 5.4 mM MgCl$_2$, 0.9 mM DTT, 1.8 mM ATP and 27 mM PIPES. The protein concentration in the actomyosin fraction was determined using the method of Bradford \cite{Bradford1976} with bovine serum albumin as the standard, and the ratios of actin or myosin to the total protein were measured using SDS-PAGE \cite{Laemmli1970} and the Fiji software package (http://fiji.sc/wiki/index.php/Fiji). Each concentration of actin and myosin was calculated relative to the concentration of total protein and the ratio of actin or myosin to the total protein. Immunoblotting was performed according to the standard protocol with anti-$\beta$-Actin (Poly6221; BioLegend) and anti-Myosin II from {\it Amoeba proteus}. Horseradish peroxidase (HRP)-conjugated secondary antibodies (Poly4053; BioLegend and Poly4064; BioLegend) were used for detection.

\subsection{Prearation and observation of the actomyosin droplets} Cell-sized water-in-oil (W/O) droplets with a lipid monolayer were prepared as described previously \cite{Kato2009,Ito2012}. In brief, the amphiphile 1,2-dioleoyl-3-trimethylammonium-propane chloride (DOTAP, Avanti Polar Lipids) was dissolved in chloroform at a concentration of 10 mM. The DOTAP solution was dried in a nitrogen stream and allowed to settle under vacuum overnight. Mineral oil (Nacarai Tesque) was added to the dried films to obtain 1 mM DOTAP, which was sonicated at 60$^\circ$C for 60 min, resulting in dispersed DOTAP in oil. Finally, to obtain cell-sized W/O droplets, 3\% (vol./vol.) actomyosin fraction was added to the oil solution, and emulsification was performed by vortexing and pipetting. To obtain transmitted light images the prepared droplets were observed using a microscope (Eclipse Ti, Nikon) at room temperature and recorded using an sCMOS camera (ORCA-Flash4.0, Hamamatsu). To reveal the distribution of actomyosin fraction in the W/O droplets, the actin filaments were stained with 10 nM Acti-stain 488 phalloidin (Cytoskeleton). Note that the phalloidin prevents actin filaments from depolymerization. Fluorescent time-lapse images were collected using a microscope (IX71, Olympus) equipped with an EM-CCD camera (iXon, Andor) and a confocal scanner unit (CSU-X1, Yokogawa). In the experiments using droplets in the absence of myosin, a mixture of 300 ${\rm \mu}$l EMPA solution and 30 ${\rm \mu}$l actin rich solution was used in place of the actomyosin fraction. Note that a myosin II inhibitor, blebbistatin, is ineffective in inhibiting myosin in {\it Amoebozoa} \cite{Limouze2004}. The obtained images were processed via Fiji and analyzed via custom routine in Igor Pro (WaveMetrics, Oregon).

\section{RESULTS AND DISCUSSION}

\subsection{Experimental results}
\begin{figure}[!ht]
 \includegraphics[bb=0 0 237 218]{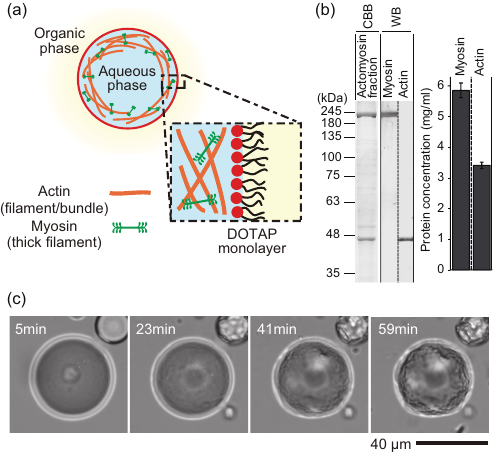}
 \caption{\label{fig:fig1} (Color online) (a) Schematic illustration of a cell-sized lipid monolayer encapsulating the actomyosin fraction. (b) Characterization of the actomyosin fraction; error bars represent standard error of the mean ($n = 5$). (c) Time development of the lipid interface induced by the contraction of the confined actomyosin fraction. The wrinkle grew on the interface. The scale bar corresponds to 40 ${\rm\mu m}$.}
 \end{figure}

In the present study, we fabricated a spherical (in its initial state) cell-sized deformable interface composed of a lipid monolayer \cite{Hase2006,Kato2009,Ito2012} that encapsulates the actomyosin fraction (Fig.~\ref{fig:fig1}(a)). As shown in Fig.~\ref{fig:fig1}(b), the actomyosin fraction used here consists of actin and myosin II extracted from {\it Amoeba proteus} \cite{Nishigami2013}. The deformation stress on the actin network cross-linked by rigor state myosins \cite{Takiguchi2011} or myosin bipolar thick filaments \cite{Nishigami2013} was exerted by the motive force of active myosin motors through the consumption of ATP, as it was confirmed by the observation that a droplet in the absence of myosin molecules did not exhibit any deformation. To establish nonspecific interconnection between the actomyosin and the interface, we employed electrostatic attraction \cite{Limozin2005} using a positively charged lipid, DOTAP. In this solution, the positively charged lipid monolayer attracts negatively charged actin filaments, accompanied with myosin thick filaments as an actomyosin complex. In this situation, we confirmed the suitable experimental conditions that actin and myosin form an actomyosin cortex structure underneath the lipid interface and cause interfacial deformation by the actomyosin contraction, as 3 mg/ml actin, 6 mg/ml myosin, and pure DOTAP lipid monolayer, by tuning the actin concentration, myosin concentration, and DOTAP fraction \cite{Nishigamiunpublished}. If the myosin concentration is higher than 6 mg/ml, e.g., 8 mg/ml, the actomyosin detached from the lipid layer and condensed to form an aggregate that was independent of the droplet boundary, as was reported in previous studies \cite{Takiguchi2011,Carvalho2013}.

\begin{figure}[!ht]
 \includegraphics[bb=0 0 227 401]{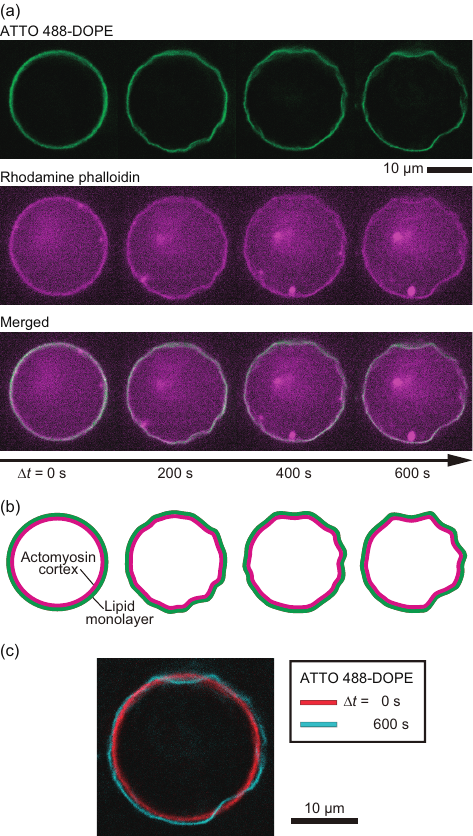}
 \caption{\label{fig:fig2} (Color online) (a) Confocal fluorescence images of lipid interface (labeled with ATTO 488-DOPE) and actin filaments (labeled with Rhodamine phalloidin) during the interfacial wrinkling. $\Delta t=0$ is defined as the time point at which the wrinkling began. Actin filaments spontaneously form a cortex structure and contract underneath the lipid interface. (b) Schematic illustrations corresponding to the confocal images shown in (a). Red and orange colors denote the lipid interface and actomyosin fraction, respectively. (c) Comparison between the original spherical interface and the wrinkled interface revealed by ATTO 488-DOPE. The original interface ($\Delta t=0\,{\rm s}$, red) is overlaid on the wrinkled interface ($\Delta t=600\,{\rm s}$, cyan). The deformation leads to the slight outward expansion of other part due to the volume conservation of aqueous phase. All scale bars correspond to 10 ${\rm \mu m}$.}
 \end{figure}

The interface of the actomyosin droplet was clearly deformed under the above-confirmed appropriate condition (Fig.~\ref{fig:fig1}(c)), which should be noted that the energy cost to deform the interface of a micrometer sized water-in-oil droplet is $10^3$--$10^4$ times higher than for an elastic lipid membrane \cite{Popescu2006,Zhou2007,Yanagisawa2013}. Nevertheless, it is obvious that the droplet oil/water interface deforms because the wrinkles revealed by the contrast in the transmitted light definitely corresponds to the deviation of the lipid layer at the oil/water interface from the initial position. This fact was further proved by the fluorescence microscopy with different dyes for the lipid interface and actomyosin as shown in Fig.~\ref{fig:fig2} (see below for the details). For a typical case shown in Fig. 1c, initially-spherical droplet exhibited interfacial wrinkling. The interfacial wrinkling was not accompanied by apparent symmetry breaking, i.e., the strong localization of actomyosin distribution, as otherwise observed in both actin gels around beads \cite{Gucht2005,Paluch2006} and reconstituted \cite{Carvalho2013,Shah2014} or cellular \cite{Charras2008,Blaser2006} actomyosin, owing to the strong adhesion between the actomyosin and the lipid interface in the present experimental condition.

Figure~\ref{fig:fig2}(a) shows a time development of the wrinkle formation visualized by fluorescently-labeled lipid interface and actin with confocal fluorescence microscopy. The actin distribution exhibits that the actomyosin inside the droplet forms cortex structure underlying the lipid interface. The same fluorescence distributions of the lipid interface and actomyosin cortex structure as shown in the merged images (Fig.~\ref{fig:fig2}(a), bottom) and its schematic illustrations (Fig.~\ref{fig:fig2}(b)) clearly verify that the interfacial wrinkling was directly induced by the underlying actomyosin cortex, which yields lateral contractile force. The comparison between the original spherical interface at the beginning of the wrinkle formation $\Delta t=0\,{\rm s}$ and the wrinkled interface at $\Delta t=600\,{\rm s}$ revealed the apparent dents of the deformed part and slight ourward expansion of the entire region at the same time, indicating the volume conservation due to the incompressibility of the inner aqueous solution (Fig.~\ref{fig:fig2}(c)).

\subsection{Characterization of actomyosin-induced deformation}
\begin{figure*}[!ht]
\centering{\includegraphics[bb=0 0 404 297]{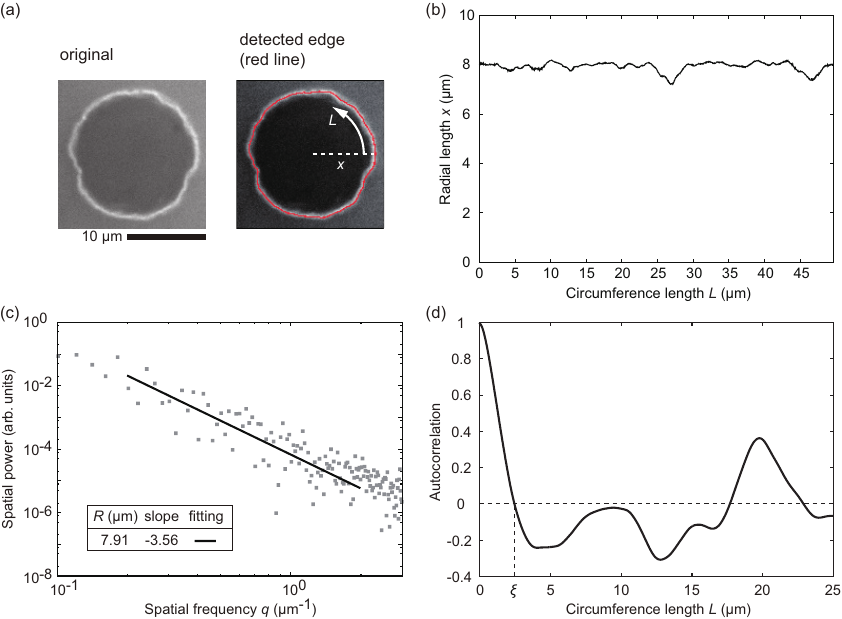}}
\caption{\label{fig:fig3} (Color online) Detailed analytical procedure of the spatial power spectrum and autocorrelation for the deformed equator shown by fluorescently-labeled actin distribution. (a) Typical example of the original fluorescence image (left) and the detected edge (right, red line). radial length $x$ is defined as the distance between two dimensional center of mass and the detected edge as a function of circumference length $L$ ranging from $0$ to $2\pi R\,{\rm (\mu m)}$. Scale bar is $10\,{\rm \mu m}$. (b) Radial length $x$ plotted versus circumference length $L$. (c) Spatial power spectrum and (d) autocorrelation calculated from the shape profile shown in ((b)). Correlation length $\xi$ is defined as the smallest length that ${\rm autocorrelation}=0$.}
\end{figure*}

To quantify and characterize the intrinsic propeties of actomyosin-induced deformation, we analyzed a wrinkled shape as a perturbative shape from the initial spherical shape. Figure~\ref{fig:fig3}(a) shows a typical example of the confocal fluorescence image of an actomyosin cortex. Based on the droplet equator revealed here, we calculated the radial position $x$ of the deformed cortex, which was taken from the two dimensional droplet center of mass, as a function of circumference length $L$ ranging from $0$ to $2\pi R$. The mean radius $R=\langle x\rangle_{\rm angle}$ was defined as the rotationally-averaged radial length. The edge detection was performed by Gaussian fittings of 1024 discrete radial intensity profiles in the vicinity of the cortex, in which the radial positions $x(L)$ were determined as the peak position of each determined Gaussian function. The red line depicted on Fig.~\ref{fig:fig3}(a) (right) indicates the detected shape from the original image (left) via this procedure. Figure~\ref{fig:fig3}(b) shows the graph of radial length $x$ versus circumference length $L$, i.e., shape profile of the deformation. From the shape profile, we characterized the actomyosin-induced deformation by calculating spatial power spectrum and autocorrelation. Gray markers shown in Figure~\ref{fig:fig3}(c) are the calculated power spectrum $\left|\frac{2}{N}\sum_{n=0}^{N-1} {x(n\Delta L)-\langle x(n\Delta L)\rangle}e^{2\pi iRqn/N}\right|^2$, where $N=1024$, $q$, and $n\Delta L=2\pi Rn/N$, are the sampling number, spatial frequency, and circumference length, respectively. The black solid line is the power law fitting, in which the exponent shown as the slope in log-log plot reflects the intrinsic properties of the deformed shape \cite{Popescu2006,Helfrich1984}. Figure~\ref{fig:fig3}(d) shows the autocorrelation $\langle x(L_0)x(L_0+n\Delta L)\rangle$, where $\langle\rangle$ denotes the average over arbitrary initial position $L_{\rm 0}$ and displacement $n\Delta L$. Because of the periodic boundary condition of the shape profile (Fig.~\ref{fig:fig3}(b)), the autocorrelation becomes negative at a certain displacement $L=n\Delta L$. We can define the correlation length $\xi$ as the smallest length of $L$ that satisfies ${\rm autocorrelation}=0$, so that $\xi$ represents the characteristic width of a concave in the wrinkled shape \cite{Sumino2011}. For example, in the case of sine function as a deviation profile from its mean value, the half-length of a convex part, i.e., $\lambda/4$ ($\lambda$; the wavelength of the sine function), appears as $\xi$ in the autocorrelation. This characteristic length of the wrinkles also reflects the underlying mechanism of the actomyosin-induced deformation.

\subsection{Curvature dependence of wrinkled shape}
\begin{figure}[!ht]
 \includegraphics[bb=0 0 191 286]{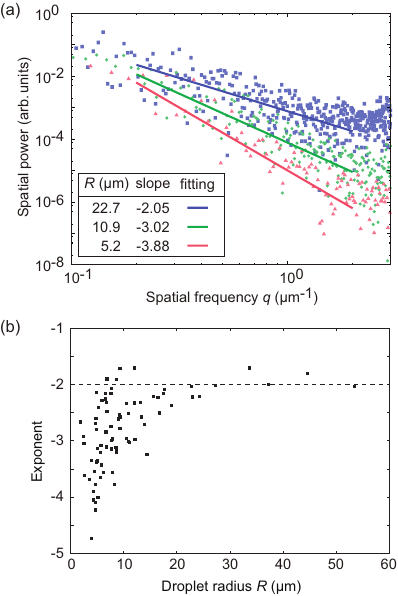}
 \caption{\label{fig:fig4} (Color online) (a) Spatial power spectra obtained from the typical equatorial shapes of various sized droplets represented by pale red, green and blue markers. Corresponding red, green and blue lines are the power fits of the experimental spectra. (b) Values of the power-law scaling exponents in the power spectra on spatial frequency with respect to droplet size. The exponents vary from $-4$ to $-2$, in correlation with the droplet size.}
 \end{figure}

We examined the spatial properties of the wrinkle emergence, which we hypothesize was caused by the lateral contractile stress generated by the actomyosin cortex underlying the lipid interface. Note that the distribution of actin observed by fluorescence was laterally homogeneous throughout the wrinkle development, except for the effect of focal depth. Here, we focused on the curvature dependence of the wrinkling deformation induced by actomyosin cortex. Along the above described procedures, the spatial power spectra were calculated for various sizes of droplets. Figure~\ref{fig:fig4}(a) shows the spatial power spectra of three typical droplets with various radii. Their power-law scaling exponents of spatial frequency $q$ change from $-4$ for smaller droplets ($R\sim5\,{\rm \mu m}$) to $-2$ for larger droplets ($R>20\,{\rm \mu m}$). Figure~\ref{fig:fig4}(b) exhibits the convergence of the scaling exponents at $-2$ with the increase in droplet size, ranging from $R=1.9\,{\rm \mu m}$ to $53.4\,{\rm \mu m}$ (Fig.~\ref{fig:fig4}(b)).

These results indicate that the dominant contributor to the deformation exhibits transitional behavior in a size dependent manner. The crossover size for the appearance of this curvature effect on actomyosin contraction is on the order of 10 ${\rm \mu m}$. The exponent $-2$ of a spatial power spectrum is generally observed in the case of interfacial tension-dominated deformation of a planar interface, whereas the exponents $-4$ and $-3$ are observed in the bending energy-dominated deformation of a planar or spherical interface \cite{Popescu2006,Helfrich1984}. Therefore, the exponents indicate that the deformed shapes of smaller and larger droplets are predominantly determined by the bending elasticity and contractile interfacial tension of the actomyosin shell, respectively. Here, the width of the concave interfaces should reflect the elastic and contractile properties of the actomyosin interacting with the cell-sized deformable interface. To gain further understanding of the non-specific interaction between the contractile actomyosin cortex and the deformable interface, we developed a theoretical model for the deformation based on the elastic energy of the actomyosin cortex. Comparison between the experimental and theoretical results should provide insight into the elastic, contractile, and geometric properties of the microscopic actomyosin cortex.  

\subsection{Theoretical model for the onset of wrinkling}
\begin{figure}[!ht]
 \includegraphics[bb=0 0 222 108]{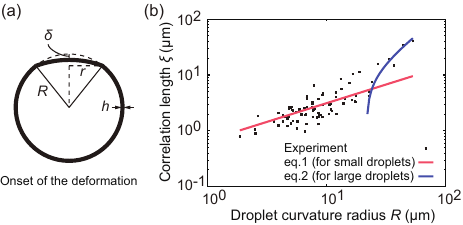}
 \caption{\label{fig:fig5} (Color online) (a) Schematic illustration of the situation considered in the model. A portion of the spherical interface is slightly dented, where $\delta$ is the depth of the dent, $r$ is the radius of the dent region, $R$ is the initial curvature radius of the droplet, and $h$ is the thickness of the actomyosin shell underlying the droplet interface. (b) Correlation length $\xi$ of the wrinkle shape profile along the circumference, plotted versus droplet size. Each correlation length is calculated at the final state of wrinkle development. Red and blue lines denote theoretical correlations between the dented radius and the droplet radius described in the main text.}
 \end{figure}

Here, we assume that actomyosin forms a homogeneous thin elastic shell-like structure with thickness $h$ and Young's modulus $E$ underneath the surface of the droplet with radius $R$. We also assume that a part of the spherical interface is slightly distorted from its original spherical shape, with the depth of distortion given as $\delta$ (Fig.~\ref{fig:fig5}(a)). The dent region stores two types of elastic energy per unit area: bending energy $f_{\rm bend} \sim Eh^3(\partial^2\delta/\partial x^2)^2 \sim Eh^3\delta^2/r^4$ and stretching energy $f_{\rm stretch} \sim Ehu^2_{ab} \sim Eh\delta^2/R^2$, where $u_{ab}$ is a component of the strain tensor of the shell. In the following, we consider the onset of the deformation from the initial spherical shape. The elastic energies over the dent region are obtained as $F_{\rm bend} \sim r^2 f_{\rm bend} \sim Eh^3\delta^2/r^2$ and $F_{\rm stretch} \sim r^2 f_{\rm stretch} \sim Eh\delta^2 r^2/R^2$. Additionally, we consider the energy cost when the other part stretches outward simultaneously with the partial dent due to the incompressibility of the inner aqueous solution (see Fig.~\ref{fig:fig2}(c)). When the local interfacial indentation occurs as shown in Fig.~\ref{fig:fig5}(a), the volume corresponding to the dented part is compensated by the slight outward expansion of the other part of the interfacial shell. Approximating that the stretching of the undeformed part compensates for the dent volume $\Delta V \sim \delta r^2$, the normal stretching length is $\Delta R \sim \delta r^2/R^2$, resulting in the additional energy cost for the distortion: $F_{\rm comp} \sim Eh(\Delta R/R)^2\cdot R^2(1+\sqrt{1-(r/R)^2}) \sim Eh\delta^2 r^4/R^4$ by neglecting the term of $\mathcal{O}((r/R)^6)$. This energy, up to the order of $(r/R)^4$, is smaller compared to $F_{\rm stretch}$ by a factor of $r^2/R^2$, but should not be neglected because the experimental ratio $\xi/R$, which corresponds to $r/R$, shown in Fig.~\ref{fig:fig5}(b) is not so much small, ranging from 0.15 to 0.83.

The interfacial deformation requires a driving force along the normal direction to the interface. In our experimental system, the driving force can be the tangential contraction provided by the contractile actomyosin shell underlying a curved interface \cite{Carvalho2013,Shah2014}. There is a general relationship between the normal stress $\sigma_{rr}$ and the tangential stress $\sigma_{\perp\perp}$ on a curved surface:
\begin{equation*}
\frac{1}{r^{\prime2}}\frac{\partial}{\partial r^\prime}(r^{\prime2} \sigma_{rr}) = \frac{2}{r^\prime} \sigma_{\perp\perp},
\end{equation*}
where $r^\prime$ is the radial coordinate \cite{Noireaux2000,Sekimoto2004,Sumino2011}. By integrating over the shell thickness $r^\prime = R \sim R + h$, the normal stress of the actomyosin shell (per unit area) can be described with the tangential stress as $\sigma_{rr} = 2/R^2\int_{R}^{R+h}dr^\prime r^\prime \sigma_{\perp\perp}  \sim 2h\sigma_{\perp\perp}/R$, where the same expression is used for the Laplace pressure. Here, the tangential stress $\sigma_{\perp\perp}$ is assumed to be a constant over the shell thickness $h$. This pressure, which originates from the tangential contraction, causes the interfacial distortion. The work done by the contraction leads to $W_{\rm contraction}\sim \sigma_{rr}\Delta V \sim h\sigma_{\perp\perp}\delta r^2/R$.

Thus, we obtain the total energy cost for the deformation:
\begin{eqnarray*}
F_{\rm total} &=& F_{\rm bend} + F_{\rm stretch} + F_{\rm comp} - W_{\rm contraction} \\ 
 		  &\sim& \frac{Eh^3\delta^2}{r^2} + \frac{Eh\delta^2 r^2}{R^2}\left(1+\frac{r^2}{R^2}\right) - \frac{h\sigma_{\perp\perp}\delta r^2}{R}.
\end{eqnarray*}
Let us consider the onset of the deformation. In our experiment, the observed dent radii maintained the same order of magnitude from the onset time to the end of the wrinkling behavior (Fig.~\ref{fig:fig2}(a)). Therefore, the initial dent radii can be regarded as the experimentally observed dent radii in the wrinkling period. Immediately after the droplet is prepared ($t = 0$ min), the tangential stress is $\sigma_{\perp\perp}(t=0)=0$. As time passes, the stress grows inside the shell; consequently, although the energy always permits a small agitation of the interface on the order of $h$, the observed macroscopic indentation of the interface begins at a certain time, $t = t_c > 0$. At this time, the signature of $F_{\rm total}$ becomes negative in the range of $0 < r < R$. According to these situations, the critical conditions for the onset of the dent formation require $F_{\rm total} = 0$ and $\partial F_{\rm total}/\partial r = 0$. These conditions produce the following correlation between the dent radius and the droplet curvature radius: 
\begin{equation}
r=(2h^2)^{\frac{1}{6}}R^{\frac{2}{3}}.
\label{eq:eq1}
\end{equation}
Here, we verified the corresponding experimental values, i.e., spatial correlation length $\xi$ of the deformed equatorial shapes for various sizes of droplets. $\xi$ was defined as the circumferential length at which the autocorrelation first becomes 0, and thus the length $\xi$ denotes the width of the concave structure along the circumference. Consequently, there is a strong positive correlation between the correlation length $\xi$ and the droplet radius $R$ (Fig.~\ref{fig:fig5}(b)). The theoretical curvature-radius dependence, indicated by Eq.~(\ref{eq:eq1}) is consistent with the experimental slope in the small-radius region with the fitting parameter $h = 218\,{\rm nm}$, which is comparable to the observable results ($< 500\,{\rm nm}$ in smaller droplets); furthermore, this value is close to the cortical thickness of living cells \cite{Charras2006,Maugis2010,Salbreux2012} and previously reconstituted systems \cite{Noireaux2000,Pontani2009,Ershov2012}. At the same time, the curve Eq.~(\ref{eq:eq1}) deviates from the experimental data in the larger-radius region of $R > 10\,{\rm \mu m}$. To elucidate the factors that influence the deviation, we return our discussion to the circumferential power spectra of the deformed shapes for various droplet sizes (Fig.~\ref{fig:fig4}(b)). The scaling exponents $-4$ and $-3$ indicate that the bending elasticity $F_{\rm bend}$ dominates the deformed shape for $R < 10\,{\rm \mu m}$, whereas the transient distribution and convergence to the exponent $-2$ indicate that the energies for area stretch (lateral tension), $F_{\rm stretch}$ and $F_{\rm comp}$, dominate the deformation for $10\,{\rm \mu m} < R < 20 \,{\rm \mu m}$ and $R > 20\,{\rm \mu m}$. Because the ratio of the bending term and tension terms is $F_{\rm bend}/F_{\rm stretch} = h^2R^2/r^4$ or $F_{\rm bend}/F_{\rm comp} = h^2R^4/r^6$, $F_{\rm bend} \ll F_{\rm stretch}$ and $F_{\rm bend} \ll F_{\rm comp}$ are realized in the large-radius region ($R > 20\,{\rm \mu m}$), where $r$ and $R$ are of the same order in experimental values (Fig.~\ref{fig:fig5}(a)). Thus, $F_{\rm bend}$ is negligible compared to the tension terms for the large-radius region. Minimizing the energies in the absence of the bending energy, we in turn obtain the following equation for the deformation of larger droplets: 
\begin{equation}
r=R^{\frac{3}{2}}\sqrt{\frac{\sigma_{\perp\perp}}{2E\delta}-\frac{1}{2R}}.
\label{eq:eq2}
\end{equation}
By fitting in the region of $R > 20\,{\rm \mu m}$ (blue line in Fig.~\ref{fig:fig5}(b)), we find $\sigma_{\perp\perp}/2E\delta \sim 0.023\,{\rm \mu m}^{-1}$. Furthermore, because $\delta$ is of the order of $1\,{\rm \mu m}$ in our experiment for $R > 20\,{\rm \mu m}$, the above relation results in the quantification of the ratio between the active force generation $\sigma_{\perp\perp}$ by actomyosin contractility and static elasticity $E$, i.e., $\sigma_{\perp\perp}/E\sim 0.01$. Note that Eq.~(\ref{eq:eq2}) is only verified under the necessary condition $\sigma_{\perp\perp}/2E\delta - 1/2R > 0$, so that the energy cost, $F_{\rm total} - F_{\rm bend}$, has a minimum value. In the case of $\sigma_{\perp\perp}/2E\delta \sim 0.023\,{\rm \mu m}^{-1}$, this necessary condition results in $R > 20\,{\rm \mu m}$, which is consistent with the situation of the larger droplets considered here.

Although the wrinkles further developed after the onset of the deformation and the dent curvature eventually became negative, the typical pattern of the wrinkled shape was determined in the early stage of the indentation. In the late stage of the deformation, the inner solution of the droplets evaporated or was ejected (data not shown) and the droplets became a bit smaller to release a part of the pressure increase as in Fig.~\ref{fig:fig1}(c), resulting in the breaking of the volume conservation inside the droplet. In addition to the contractile force of the actomyosin shell, such increase in area-to-volume ratio in the late stage could induce complex buckling of the elastic shell as well. While the wrinkled width $\xi$ or $r$ was determined in the early stage considered in the model, to investigate the complex deformation mechanism in the late stage in detail would be a future work for this system.

\section{CONCLUSIONS AND PROSPECTS}
Reconstituted actomyosin cortex underlying a deformable lipid interface successfully caused the wrinkling deformation of the interface for the first time. Furthermore, the shape analysis and theoretical descriptions revealed the contributions of the bending elasticity, stretching elasticity, and contractility of the reconstituted cortex structure on the interfacial deformation, which anomalously depends on the cell-sized droplet radius, i.e., the interfacial curvature. In other words, we demonstrated the physical basis that even the primitive geometrical coupling between the deformable interface and confined actomyosin exhibits rich variety of interfacial deformability. The geometrical origin in our findings should be relevant and taken into consideration for not only the purified systems but also living systems such as the contraction of the cell membrane with the contractile cell cortex at the posterior during the amoeboid motion of a cell. To study the applicability of the present curvature dependence of the contractile cortex in living cells such as {\it Amoeba proteus} would be one of the most interesting future works. With respect to the objective of reconstituting cellular motility, the spatiotemporal regulation of the structural formations should be addressed as the next step of this work by exploring the implementation of the frameworks of actin-related molecules and their signaling networks. Our present findings on how actomyosin behaves within a cell-sized compartment with a deformable interface provides an experimental basis for this purpose, and it takes an important step toward artificial cellular deformation in the step-wise reconstitution of the living state. Together, these investigations may lead to the further clarification of complex biological phenomena involving geometrically coupled actin and myosin functions.


\begin{acknowledgments}
We thank N. Yoshinaga and K. Takiguchi for fruitful discussions and comments on the manuscript. This work is supported by JSPS KAKENHI Grant Number 26707020; by MEXT KAKENHI Grant Numbers 25103012, 26115709 (ea. M.I.) and 25117520 (S.S.); by a Grant-in-Aid for JSPS Fellows (No. 25-1297, H.I.); and by a BioLegend/Tomy Digital Biology Research Grant 2013 (Y.N.).
\end{acknowledgments}

\appendix*
\section{EFFECT OF INTERFACIAL TENSION}
In the main text, we only considered the elasticity and contractility of actomyosin shell. However, the surface tension of the lipid interface may also contribute to the deformation energy cost, because the lipid interface was also deformed to the same shape with the shell, as shown in Figure \ref{fig:fig2}(a). In the same theoretical framework used in the main text, the contribution of surface tension $\gamma$ for the dent region can be described as the work done by Laplace pressure
\begin{equation}
\frac{2\gamma}{R}\Delta V \simeq -\frac{2\pi\gamma\delta r^2}{R},
\end{equation}
where $\Delta V\simeq\pi\delta r^2$ is the decreased volume by the dent defined as negative. On the other hand, the other region should slightly expand outward due to the imcompressibility of the inner aqueous phase, resulting in the radial expansion $\Delta R\sim\frac{\delta r^2}{R^2}$. This contributes to the interfacial energy cost:
\begin{equation}
\gamma\Delta A\simeq\frac{8\pi\gamma\delta r^2}{R},
\end{equation}
where $\Delta A\simeq 4\pi\left\{(R+\Delta R)^2-R^2\right\}$ is the approximated surface area difference of the droplet before and after the outward expansion. Therefore, the interfacial energy cost of the deformation becomes $F_{\rm int}\simeq\frac{8\pi\gamma\delta r^2}{R}-\frac{2\pi\gamma\delta r^2}{R}\sim\frac{\gamma\delta r^2}{R}$. The total energy cost including the elasticity and contractility of actomyosin shell is then obtained by replacing the lateral stress $\sigma_{\perp\perp}$ by effective lateral stress $\sigma_{\perp\perp}-\frac{\gamma}{h}$:
\begin{eqnarray*}
F_{\rm total} &=& F_{\rm bend} + F_{\rm stretch} + F_{\rm comp} - W_{\rm contraction} + F_{\rm int}\\ 
 		  &\sim& \frac{Eh^3\delta^2}{r^2} + \frac{Eh\delta^2 r^2}{R^2}\left(1+\frac{r^2}{R^2}\right) - \frac{h\delta r^2}{R}\left(\sigma_{\perp\perp}-\frac{\gamma}{h}\right).
\end{eqnarray*}
The same fitting in the main text reads $\frac{\sigma_{\perp\perp}-\frac{\gamma}{h}}{E}\sim0.01$. Thus, the experimentally observed wrinkling is realized in the case that the contraction exceeds the interfacial tension, namely $\sigma_{\perp\perp}>\frac{\gamma}{h}$. Our experimental result indicates the lower limit of the ratio between the active contractility $\sigma_{\perp\perp}$ and static elasticity $E$ as $\frac{\sigma_{\perp\perp}}{E}>0.01$. Note that $h\sim200\,{\rm nm}$ and $\gamma\sim10\,{\rm mN/m}$ result in $\frac{\gamma}{h}\sim1\,{\rm nN/m^2}$.

\bibliography{reference.bib}

\end{document}